\scrollmode

\documentclass[a4paper,12pt,prb]{revtex4}
\usepackage{graphicx}
\usepackage{amssymb}
\usepackage{url}

\begin{document}

\title{Self-consistent Ornstein-Zernike approximation for molecules with
soft
cores}

\author{J.~S.~H{\o}ye$^1$ and A.~Reiner$^2$,\\
Teoretisk fysikk, Institutt for fysikk,\\
Norges teknisk-naturvitenskapelige universitet (NTNU) Trondheim,\\
H\o gskoleringen 5, N-7491 Trondheim, Norway.\\
{}$^1$e-mail: {\tt Johan.Hoye@phys.ntnu.no}\quad
{}$^2$e-mail: {\tt areiner@tph.tuwien.ac.at}}

\begin{abstract}
The Self-Consistent Ornstein-Zernike Approximation (SCOZA) is an
accurate liquid state theory.  So far it has been tied to interactions
composed of hard core repulsion and long-range attraction, whereas
real molecules have soft core repulsion at short distances.  In the
present work, this is taken into account through the introduction of
an effective hard core with a diameter that depends upon temperature
only.  It is found that the contribution to the configurational
internal energy due to the repulsive reference fluid is of prime
importance and must be included in the thermodynamic self-consistency
requirement on which SCOZA is based.  An approximate but accurate
evaluation of this contribution relies on the virial theorem to gauge
the amplitude of the pair distribution function close to the molecular
surface.  Finally, the SCOZA equation is transformed by which the
problem is reformulated in terms of the usual SCOZA with fixed hard
core reference system and temperature-dependent interaction.\end{abstract}

\maketitle

\newpage

\section{Introduction}

The Self-Consistent Ornstein-Zernike Approximation (SCOZA) has been
found to give very accurate results for the equation of state for
fluids and lattice gases.  This approximation was proposed by H{\o}ye\
and Stell \cite{scoza:12,scoza:11}.  Some preliminary results were
obtained for supercritical temperatures \cite{scoza:11,scoza:9}.

However, for subcritical temperatures there were numerical problems
connected to the singular behavior along the spinodal curve and the
no-solution region inside it.  The first successful subcritical
solution was made by Dickman and Stell for the Ising model or the
lattice gase case \cite{scoza:1}.  It turned out that results were
very accurate.

Borge and H{\o}ye\ made a more general numerical investigation of the
SCOZA equation of state in the critical region \cite{scoza:20}.
Clearly scaling was not fulfilled very close to the critical point.
But apart from that the critical behavior was close to that of real
fluids.  Especially it was noted that the critical exponent $\beta$
for the curve of coexistence was equal or close to 0.35.  This value
was subsequently confirmed by numerical evaluations by others
\cite{pini:stell:dickman:1998}.

Also Borge and H{\o}ye\ compared SCOZA results with experimental data
on $\mathrm{CO_2}$ in the critical region \cite{CO2:data}.  By closer
study of these data on a scaling plot it was realized that they did
not collapse onto a single scaling curve but gave a series of
close-lying curves instead, and the SCOZA results were consistent with
such deviations from scaling \cite{scoza:20}.

H{\o}ye, Pini and Stell then made a closer investigation of the
critical region of SCOZA\cite{scoza:25}.  Their analysis showed that
SCOZA fulfills a generalized kind of scaling instead of the usual one,
and they obtained the value 0.35 for the critical exponent $\beta$.

SCOZA was then applied to continuum fluids, and Pini, Stell and
H{\o}ye\ considered the hard sphere fluid with interaction of Yukawa
form \cite{pini:stell:hoye:1998}.  The results aggreed well with
simulation data.  Furthermore, interactions using more Yukawa terms
were also considered \cite{scoza:15}.  This allowed approximations to
the Lennard-Jones (LJ) potential of realistic fluids to be studied
\cite{scoza:15}.  Accurate results were also obtained for the more
short-ranged fullerene interactions \cite{scoza:24,scoza:29}.

So far SCOZA has been applied to continuum systems with hard cores
perturbed by an attractive interaction only.  Real molecules, on the
other hand, have soft cores.  In order to obtain more accurate
results, this should therefore be taken into account.  To do so we use
hard spheres with an effective diameter $d$.  In combination with the
self-consistency requirement of SCOZA, however, the use of an
effective diameter turns out to become a non-trivial problem.  This is
connected to the necessity of properly taking into account the
contribution to the internal energy due to the repulsive interaction.
Furthermore, a numerical procedure using the semi-analytical solution
of the Ornstein-Zernike (OZ) equation becomes more problematic as a
grid of fixed densities $\rho$ implies varying packing fractions
$\eta=\frac\pi6\rho d^3$ when the effective diameter $d$ changes.
While a fully numerical solution of the OZ\ relation is possible
\cite{scoza:26} and sidesteps this particular problem, it entails a
huge increase in computational complexity and cost.

In the present work, on the other hand, we avoid the need for heavy
numerics.  Instead we opt for a more conceptual approach to the
introduction of soft cores into SCOZA by means of effective hard
cores.  In doing so we prefer here a prescription for the hard core
diameter $d$ that depends only upon temperature
(section~\ref{sec:effective}).  It turns out that an accurate
evaluation of the reference system contribution to the configurational
internal energy plays an important role.  This quantity can be linked
to the pressure and hence to the virial theorem
(section~\ref{sec:virial}) and must be taken into account for the
energy route part of the SCOZA self-consistency problem
(section~\ref{sec:scoza}).  A further refinement uses the virial
theorem once more, this time to gauge the amplitude of the pair
distribution function close to the core (section~\ref{sec:contact}).
We conclude with a formal mapping of the soft-core SCOZA for fixed
interaction onto the usual SCOZA with fixed hard core and a
temperature dependent potential.  This essentially eliminates any
softness and so provides a way of avoiding the complexities associated
with it (section~\ref{sec:unit}).

%22222222222222222222222222222222222222222222222222222222222222222222222
\section{Interaction and effective hard cores}

\label{sec:effective}

The pair interaction for real atoms that are neutral is typically
given by the LJ\ interaction
\begin{displaymath}
\phi(r) = 4\epsilon\,\left[\left(\frac\sigma r\right)^{12}
-\left(\frac\sigma r\right)^6\right],
\end{displaymath}
with length and energy scales of $\sigma$ and $\epsilon$,
respectively.  (From now on we will use units where
$\sigma=\epsilon=1$.)  In approximations, the repulsive part of this
interaction where $\phi(r)>0$ is commonly replaced by a hard core.
This has also been done in earlier SCOZA computations where the hard
core diameter $d$ was kept fixed.

State-dependent effective hard cores have been used earlier to
evaluate the equilibrium properties of fluids, and there exist various
recipes; see Ref.~\onlinecite{bas:reformulation} for a recent
compilation and comparison in the context of thermodynamic
perturbation theory.  The most simple schemes such as the
Barker-Henderson prescription \cite{barker:henderson} use an effective
diameter $d$ that depends on temperature only, whereas more advanced
recipes include a density dependence, too \cite{hansen:mcdonald}.

In the present work we limit ourselves to a $d$ that depends only upon
temperature.  The main reason for this is the added complexity of the
problem when combining it with the self-consistency of SCOZA.  In
particular, the latter requires that the reference system internal
energy becomes part of the thermodynamic self-consistency and must be
properly accounted for.  This inclusion of the contribution of the
soft cores in the self-consistency problem is the fundamental
difference between this method and another one which is a combination
of SCOZA and a first order perturbation theory proposed as a way of
handling soft repulsive cores\cite{allg:27} and studied
recently\cite{ar:20}.

Another reason for the use of a $d$ that depends only upon temperature
is the connection to the hard sphere equation of state that becomes
unique.  In other words, both the virial theorem and the
compressibility theorem will remain consistent if they are consistent
for the reference system with temperature independent~$d$.

The importance of $d$ for quantitative accuracy is seen from the fact
that the mean-field critical temperature and density vary with $d$ as
$d^{3}$ and $d^{-3}$ respectively when the interaction $\psi(r)$ is
held constant.  From the second virial coefficient of the repulsive
part of the LJ\ interaction one will find roughly $d^3\sim0.92$ at the
critical temperature if $\phi(r)$ for $r<1$ is replaced by an
effective hard core. This is in satisfactory agreement with the SCOZA
result with fixed $d$ when compared with simulations. With temperature
independent $d=1$ SCOZA yields the value $T_c = 1.245$ with a 3-Yukawa
fit to the Lennard-Jones interaction for $r>1$, \textit{cf.} Fig.~4.3
of Ref.~\onlinecite{paschinger:dr}. For simulations performed for the
full LJ interaction, one has found $T_c=1.310$ \cite{lj:3} and 1.313
\cite{gubbins} by molecular dynamics, and $ T_c=1.3120(7)$ \cite{lj:2}
and 1.326(2) \cite{caillol} by Monte Carlo methods.  Previous SCOZA
evaluations but with temperature independent $d=1$, however, yielded
values $ T_c =1.304$, 1.293 \cite{scoza:15}, and 1.305
\cite{paschinger:dr} of the critical temperature.  But these values
are based on 2-Yukawa fits to the LJ interaction with a compensation
for the soft core such that the right second virial coefficient is
reproduced. The variations in these latter numbers is due to the
precise potential fit and the detailed form of the direct correlation
function used with respect to the reference system hard cores.
   
Without thermodynamic self-consistency, results for isotherms at
different temperatures are independent.  The effective $d$ for a given
temperature then determines the equation of state for that
temperature.  This is the case for common fluid theories.  For SCOZA,
on the other hand, this is no longer true.  But away from the critical
region this coupling cannot be important as other theories are
accurate there, too.  One approximate way to implement SCOZA with
temperature-dependent $d$ can then be to solve the equations with the
same $d$ at all temperatures but to use the results only for the
isotherm corresponding to the chosen $d$.  Repeated evaluations with
different $d$ will then give the full phase diagram.  This approach
bears some resemblance to the SCOZA-based perturbation theory
mentioned above \cite{ar:20}.  Just as the latter, however, such a
procedure is only approximate, and compiling a full phase diagram
requires substantial computer resources far in excess of those needed
for a single run of the SCOZA program.

The division of the pair interaction $\phi(r)$ into a reference system
part $\phi_0(r)$ and a perturbing part $\psi(r)$ can be performed in
various ways.  As suggested above, one can use the repulsive part
where $\phi(r)>0$ as the reference system interaction
\cite{barker:henderson},
\begin{displaymath}
\phi_0(r) = \left\{
\begin{array}{ll}
\displaystyle \phi(r)&\displaystyle r<1\\
\displaystyle 0&\displaystyle r>1.
\end{array}
\right.
\end{displaymath}
With this the perturbing interaction is
\begin{equation}\label{eq:split:1}
\psi(r) =  \left\{
\begin{array}{ll}
\displaystyle 0&\displaystyle r<1\\
\displaystyle \phi(r)&\displaystyle r>1.
\end{array}
\right.
\end{equation}
However, there are more choices for this splitting of the interaction,
and an alternative is to use \cite{wca}
($r_\mathrm{m}=2^\frac16$)
\begin{displaymath}
\phi_0(r) = \left\{
\begin{array}{ll}
\displaystyle \phi(r)-\phi(r_\mathrm{m})&\displaystyle r<r_\mathrm{m}\\
\displaystyle 0&\displaystyle r>1
\end{array}
\right.
\end{displaymath}
\begin{displaymath}
\psi(r) =  \left\{
\begin{array}{ll}
\displaystyle \phi(r_\mathrm{m})&\displaystyle r<r_\mathrm{m}\\
\displaystyle \phi(r)&\displaystyle r>r_\mathrm{m}.
\end{array}
\right.
\end{displaymath}
As was verified by H{\o}ye\ and Borge, SCOZA requires a perturbing
interaction that is mainly attractive, or else the equations cannot be
solved \cite{hoye:borge:1998}.  Clearly, the above suggested
splittings of the potential fulfill this condition.

Furthermore, in order to keep the numerical implementation less
demanding it is desirable to approximate the interaction outside the
effective hard core as a sum of Yukawa terms.  This allows the OZ\
equation to be solved in a semi-analytic way.  For a potential like
the LJ\ one, such a multi-Yukawa form of $\phi(r)$ outside the core is
easily found. For example, a simple non-linear least-squares fit of a
sum of three Yukawa terms constrained to reproduce $\phi(1)=0$
converges rapidly and gives a result that is essentially
indistinguishable from the original LJ form for
$r>1$\cite{paschinger:dr}.  When $d<1$, however, this fit is certainly
not constant for $d < r < 1$ as mandated by the two types of splitting
mentioned above.  For instance one can then add one more Yukawa term
to approximate the desired form; this will generally be a rather
short-ranged function that hardly contributes beyond $r=1$.  For
prescription~(\ref{eq:split:1}) both the range and the amplitude of
the additional Yukawa term can be fixed by imposing, \textit{e.~g.{}},
$\psi(d) = \psi'(d) = 0$.  At any rate, any remainder of $\psi(r)$ not
accounted for by this sum of Yukawa terms is added to $\phi_0(r)
\equiv \phi(r)-\psi(r)$ and so enters the computation through the
evaluation of the effective diameter.

In solving SCOZA, the reference system is used as a boundary condition
at temperature $T\to\infty$, or $\beta=0$ where $\beta = 1/(k_B\,T)$
and $k_B$ is Boltzmann's constant.  Strictly speaking, the reference
system becomes the ideal gas in this limit for a soft core,
\textit{i.~e.{}}, $d\to0$ which is far from unity. However, use of the
ideal gas as reference system may give rise to additional problems,
especially in the numerical implementation.  So we have not tried to
investigate this possibility further.  Instead we have focused upon
the situation with an effective hard core diameter $d$ near 1.  The
justification for this lies in our arguments above. They mean that
different temperatures and densities do not couple significantly away
from the critical point anyway.  We can therefore start at $\beta=0$
with a non-vanishing value of $d$ corresponding to, say, twice the
critical temperature, and start to vary $d$ only at lower temperatures
for which SCOZA values then will be valid.

%33333333333333333333333333333333333333333333333333333333333333333333333333333333333
\section{Reference system and repulsive internal energy}

\label{sec:virial}

In approximating a soft repulsive core by an effective hard core, the
simplest prescription is to let the diameter $d$ depend only upon
temperature.  One possibility is to define $d$ such that the second
virial coefficient for the soft particles coincides with that of the
effective hard cores,
\begin{equation} \label{eq:d:def}
\frac{4\pi}3\,d^3 = \int\left(1-e^{-\beta\phi_0(r)}\right)\,{{\mathrm
d}}\vec r.
\end{equation}
The prescription of Ref.~\onlinecite{barker:henderson} is slightly
different from this in that it replaces the three-dimensional
integral~(\ref{eq:d:def}) with a one-dimensional one.  However, the
result will be the same to leading order in the difference $1-d$ which
is considered small. In this connection it can be mentioned that the
precise prescription is not crucial since both give a $d$ that depends
only upon $\beta$, and the SCOZA problem needs only $d(\beta)$ as
input, not its prescription.

At high density there are better approximations with density
dependence, but we expect the deviations of such choices from
Eq.~(\ref{eq:d:def}) to be small when the soft potential can be
considered steep at the molecular surface.

In contrast to strict hard spheres, a soft repulsive potential implies
that the reference system also contributes to the configurational
internal energy.  This can be related to the pressure and ultimately
to the virial theorem.  The equation of state can be written as
\begin{equation} \label{eq:ref:eos:i0}
\beta p = \rho + \left(1-\rho\frac\partial{\partial\rho}\right)\,I_0,
\end{equation}
where $p$ is the pressure, $I_0 = - \beta\rho f_0$, and $f_0$ is the
excess (beyond the ideal gas) Helmholtz free energy per particle.  For
effective hard cores $I_0/\rho$ only depends on the packing fraction
$\eta$,
\begin{displaymath}
I_0 = \rho\,y(\eta).
\end{displaymath}
Insertion into Eq.~(\ref{eq:ref:eos:i0}) thus gives
\begin{equation} \label{eq:ref:df:deta}
\beta p = \rho \left(1-\eta\frac{\partial y}{\partial\eta}\right).
\end{equation}
The configurational or excess internal energy $u_0$ per particle of
the reference system is now
\begin{displaymath}
\begin{array}{rl}
\displaystyle \rho u_0 &\displaystyle = -\frac{\partial I_0}{\partial\beta}
= -\rho \frac{\partial y}{\partial\eta}
\frac{\partial \eta}{\partial\beta}\\
\displaystyle   \\
\displaystyle &\displaystyle =-\rho\eta\frac{\partial y}{\partial\eta}
\frac{\partial \ln d^3}{\partial\beta}
\end{array}
\end{displaymath}
and finally, using Eq.~(\ref{eq:ref:df:deta}) for ${\partial
y}/{\partial\eta}$,
\begin{displaymath}
\rho u_0 = (\beta p-\rho)\frac{\partial \ln d^3}{\partial\beta}.
\end{displaymath}

The virial theorem for hard spheres implies
\begin{displaymath}
\beta p - \rho = \frac{2\pi}3 d^3 \rho^2 n_0(d+)
\end{displaymath}
and so connects the above internal energy to the contact value
$n_0(d+)$ of the pair distribution function,
\begin{equation} \label{eq:ref:u0}
\rho u_0 =
\frac{2\pi}3 d^3 \rho^2 n_0(d+)
\frac{\partial \ln d^3}{\partial\beta}.
\end{equation}

To gain some intuition for the consistency of this result, we can
insert the definition~(\ref{eq:d:def}) for the effective diameter.
This yields
\begin{displaymath}
\frac{\partial \ln d^3}{\partial\beta}
= \frac1{d^3}\frac{\partial d^3}{\partial\beta}
= \frac{\int\phi_0(r)e^{-\beta\phi_0(r)}{{\mathrm d}}\vec
r}{\frac{4\pi}3\,d^3}
\end{displaymath}
and thus finally
\begin{displaymath}
\rho u_0 = \frac12 \rho^2 n_0(d+)\int\phi_0(r)e^{-\beta\phi_0(r)}{{\mathrm
d}}\vec r.
\end{displaymath}
This is the low density value of the internal energy, amplified by the
contact value $n_0(d+)$ for higher densities.  Compared with the exact
$u_0$, the pair distribution function is here approximated by
$n_0(d+)\*e^{-\beta\phi_0(r)}$.  The accuracy of this increases as the
repulsive part of the potential becomes less soft. A different
prescription for $d(\beta)$ such as that of
Ref.~\onlinecite{barker:henderson} gives different expressions for the
above two integrals.  But otherwise the precise choice of $d(\beta)$
is of no consequence for the remainder of this work.

%4444444444444444444444444444444444444444444444444444444444444444
\section{SCOZA equations}

\label{sec:scoza}

The SCOZA approach is based upon thermodynamic consistency between the
energy and compressibility routes to thermodynamics.  These routes are
connected \textit{via}\ the thermodynamic relation
\begin{equation} \label{eq:scoza:pde}
\frac{\partial a}{\partial \beta}
= \rho\frac{\partial^2\rho u_t}{\partial\rho^2},
\end{equation}
where
\begin{displaymath}
a = \frac{\partial\beta p}{\partial\rho}
\end{displaymath}
is the reduced inverse compressibility and $u_t$ is the total
configurational internal energy per particle.  Both $a$ and $u_t$ are
evaluated from the pair correlation function in different ways,
\textit{viz.{}}, by the compressibility and energy routes to
thermodynamics.  In general, the pair structure is known only
approximately, and $a$ and $u_t$ give different thermodynamics.  In
SCOZA, on the other hand, consistency between the two routes in the
form of Eq.~(\ref{eq:scoza:pde}) is enforced by adjusting an unknown
parameter, usually the amplitude of the direct correlation function
$c(r)$ outside the hard core.  Specifically, in the Mean Spherical
Approximation (MSA) the contribution to $c(r)$ from the perturbing
attractive interaction $\psi(r)$ is
\begin{equation} \label{eq:attr:c}
c(r) = -\beta\psi(r),\qquad r>d.
\end{equation}
SCOZA replaces $\beta$ in the above relation by an effective value
that depends on both temperature and density and is obtained from the
solution of Eq.~(\ref{eq:scoza:pde}).  The total correlation function
$h(r) = n(r)-1$ is then obtained from the core condition
\begin{displaymath}
h(r) = -1, \qquad r < d,
\end{displaymath}
and the OZ\  equation
\begin{displaymath}
\tilde h(k) = \tilde c(k) + \rho\tilde c(k)\tilde h(k).
\end{displaymath}
(The tilde marks Fourier transforms.)  One can then evaluate $a$ as
well as $u$, the internal energy contribution from the attractive
interaction alone, according to
\begin{displaymath}
a = 1-\rho \tilde c(0),
\end{displaymath}
\begin{equation} \label{eq:attr:u}
u = \frac12 \rho \int\psi(r)(h(r)+1){{\mathrm d}}\vec r.
\end{equation}
In this way, both $a$ and $u$ are in principle obtained as functions
of the effective temperature $\beta_e$ for any given density and
interaction.  Consequently, $u$ (or $a$) can replace $\beta_e$ as the
free parameter so that correspondingly $a$ (or $u$) can be expressed
in terms of the former.  Inserting this into Eq.~(\ref{eq:scoza:pde})
one then obtains the SCOZA partial differential equation (PDE) for $u$
(or $a$).

%55555555555555555555555555555555555555555555555555555555555555555555555555555
\section{Contact value of the pair distribution function}

\label{sec:contact}

For soft repulsive interactions the total internal energy $u_t$ to be
used in Eq.~(\ref{eq:scoza:pde}) is the sum of the
contributions~(\ref{eq:ref:u0}) and~(\ref{eq:attr:u}),
\textit{i.~e.{}},
\begin{equation} \label{eq:u:sum}
u_t = u_0 + u.
\end{equation}
At first sight one might expect $u_0$ to play a minor role.  But a
closer investigation shows that it is crucial for obtaining results
consistent with a changing diameter $d$.  The reason for this is that
the reference system enters primarily as the boundary condition of the
PDE\ at $\beta=0$.  For $\beta>0$, a change in $d$ is not ``seen''
except through $u_0$ (and a small perturbation of $u$ due to small
changes in $h(r)$).  Neglecting $u_0$ thus means essentially keeping
$d$ fixed at its $\beta=0$ value.  Our preliminary numerical work
strongly indicated this importance of the $u_0$ term.

Since $u_0$ plays such an important role, the accuracy of the contact
value $n(d+) = 1+h(d+)$ becomes of interest.  For $\beta>0$, $n(d+)$
deviates from $n_0(d+)$, and clearly the former more accurately
describes the energy due to the reference system interaction
$\phi_0(r)$.  With a direct correlation function of the
form~(\ref{eq:attr:c}), however, there is no reason to expect that the
$h(r)$ obtained is accurate close to the hard core.  In particular,
the contact value $h(d+)$ strongly depends on the choice of $\psi(r)$
at $r=d$ and thus on the potential fit with the added Yukawa tails.  A
more reliable method of obtaining the contact value is thus desirable.

One appealing possibility is provided by the virial theorem that then
also, to a certain degree, enters the SCOZA where it has played no
role traditionally.  H{\o}ye\ and Stell earlier proposed full
consistency between the energy, virial and compressibility routes
\cite{scoza:11}, but this requires structure functions depending on
two free parameters and has not been considered numerically so far,
nor will it be considered here. But on the other hand we can still
obtain desired information about the contact value $n(d+)$ via the
virial theorem using the SCOZA quantities as input.

{}From the virial theorem for hard cores with attraction, the contact
value $n(d+)$ follows as
\begin{displaymath}
\beta p = \rho + \frac{4\pi}6d^3\rho^2n(d+) + \beta\rho(u-v),
\end{displaymath}
where the virial integral is split into two parts, with $u$ given by
Eq.~(\ref{eq:attr:u}).  The expression for the remainder $v$ then
becomes
\begin{displaymath}
\begin{array}{rl}
\displaystyle v &\displaystyle = u + \frac16\rho\int(\vec
r\nabla\psi(r))n(r){{\mathrm d}}\vec r\\
\displaystyle &\displaystyle =\frac16\rho\int\nabla(\vec
r\psi(r))n(r){{\mathrm d}}\vec r,
\end{array}
\end{displaymath}
with the integrations restricted to $r>d$.  Note that $v$ vanishes in
the mean field limit as then $n(r)=1$.  Expression~(\ref{eq:ref:u0})
for $u_0$ can now be expressed in terms of the new contact value
$n(d+)$ instead of the reference system value $n_0(d+)$.  With the
above relations we find
\begin{equation} \label{eq:ref:u0:virial}
\rho u_0
= (\beta p -\rho - \beta\rho u + \beta\rho v) 
\frac{\partial \ln d^3}{\partial\beta}.
\end{equation}
Thus the additional complication of evaluating $u_0$ is the evaluation
of the integral for $v$.

%666666666666666666666666666666666666666666666666666666666666666666666666666
\section{Transformation to unit diameter}

\label{sec:unit}

With varying $d$ there is an additional problem if the analytic
solution of the OZ\ equation for a sum of Yukawa terms is used.  The
latter provides $a$ as a function of $u$ only at constant packing
fraction and interaction whereas the PDE\ requires the temperature
derivative to be taken at constant density.  When $\beta$ and thus $d$
change while the density grid in the discretization is kept fixed, the
packing fraction also changes and the relation between $a$ and $u$
becomes less direct and more challenging to evaluate numerically.

If now the varying $d$ problem can be transformed into a situation of
fixed unit diameter $d=1$, the discrepancy mentioned will not arise
and application of the analytic solution of the OZ\ equation will be
simpler.  The price to pay is an interaction that varies with $\beta$
in the transformed problem.  As we will see below, this works out
nicely and gives equations that can be given a direct physical
interpretation.

To obtain the desired transformation we introduce a number of
quantities, marking those of the unit diameter problem by a
subscript~1:
\begin{displaymath}
\begin{array}{ccc}
\displaystyle \rho_1=\rho d^3,
&\displaystyle \beta_1=\beta/d^3,
&\displaystyle p_1=pd^6,\\
\displaystyle u_1=ud^6,
&\displaystyle v_1=vd^3,
&\displaystyle \phi_1(r_1) = d^3\phi(r),\\
\displaystyle r_1=r/d,
&\displaystyle {{\mathrm d}}\vec r_1={{\mathrm d}}\vec r/d^3,
&\displaystyle n_1(r_1) = n(r).
\end{array}
\end{displaymath}
With Eqs.~(\ref{eq:scoza:pde}), (\ref{eq:u:sum}), and
(\ref{eq:ref:u0:virial}) we
have the SCOZA PDE
\begin{displaymath}
\frac{\partial a}{\partial\beta}
= \rho \frac{\partial^2}{\partial\rho^2}\left[
\rho u + A (p-\rho u+\rho v)\right],
\end{displaymath}
where for brevity we put
\begin{displaymath}
A = \beta \frac{\partial\ln d^3}{\partial\beta}
= \frac\beta{d^3}\frac{\partial d^3}{\partial\beta}.
\end{displaymath}
For a $d$ that depends only on $\beta$, the introduction of the new
quantities on the right hand side is done in a straightforward way.
With $\partial^2/\partial\rho^2 = d^6 \partial^2/\partial\rho_1^2$ we
get
\begin{equation} \label{eq:jh:27}
\frac{\partial a}{\partial\beta}
= \frac1{d^3} \rho_1 \frac{\partial^2}{\partial\rho_1^2}\left[
\rho_1 u_1 + A (p_1-\rho_1 u_1+\rho_1 v_1)\right].
\end{equation}
For $a$ we find likewise
\begin{displaymath}
a = \frac{\partial(\beta p)}{\partial\rho}
= \frac{\partial(\beta_1 p_1)}{\partial\rho_1}
= a_1.
\end{displaymath}
The derivative with respect to $\beta$ then becomes
\begin{displaymath}
\begin{array}{rl}
\displaystyle \frac{\partial a}{\partial\beta}
&\displaystyle =\frac{\partial\beta_1}{\partial\beta}
\frac{\partial a_1}{\partial\beta_1}
+\frac{\partial\rho_1}{\partial\beta}
\frac{\partial a_1}{\partial\rho_1}\\
\displaystyle &\displaystyle =\left(\frac1{d^3}-\frac\beta{d^6}
\frac{\partial d^3}{\partial\beta}\right)
\frac{\partial a_1}{\partial\beta_1}
+\rho d^3\frac1\beta A\frac{\partial a_1}{\partial\rho_1}\\
\displaystyle &\displaystyle =\frac1{d^3}\left[(1-A)\frac{\partial
a_1}{\partial\beta_1}
+\rho_1A\frac{\partial^2p_1}{\partial\rho_1^2}
\right].
\end{array}
\end{displaymath}
Upon insertion into Eq.~(\ref{eq:jh:27}) the $p_1$ terms cancel and we
obtain
\begin{equation} \label{eq:scoza:transformed}
\frac{\partial a_1}{\partial\beta_1}
= \rho_1 \frac{\partial^2}{\partial\rho_1^2}\left[
\rho_1 u_1 + A_1 \rho_1 v_1\right],
\end{equation}
where
\begin{displaymath}
A_1 = \beta_1 \frac{\partial\ln d^3}{\partial\beta_1}
=\frac A{1-A}.
\end{displaymath}
The latter equality follows from
\begin{displaymath}
\left(\frac{1-A}{d^3}\right)^{-1}
= \frac{\partial\beta}{\partial\beta_1}
= d^3 + \beta\frac{\partial d^3}{\partial\beta_1}
= d^3(1+A_1).
\end{displaymath}

The resulting Eq.~(\ref{eq:scoza:transformed}) can be given a direct
physical interpretation in terms of the transformed system.  The
latter consists of hard spheres of fixed diameter ($d_1=1$), where now
there is an attractive pair interaction that depends upon the
temperature, $\phi_1(r_1) = d^3\phi(r) = d^3\phi(r_1 d)$.
Furthermore, the soft repulsive interaction is no longer present as
the transformed system has hard cores.  In the usual virial graph
expansion, the Helmholtz free energy per particle at given density
will depend only upon the interaction as before (besides temperature
and density).  This will remain the same even though $\phi_1(r_1)$
depends on the temperature.  The total configurational internal energy
per particle obtained from the pair distribution function is thus
\begin{displaymath}
\begin{array}{rl}
\displaystyle u_{1t}
&\displaystyle =\frac{\partial(\beta_1f_1)}{\partial\beta_1}\\
\displaystyle &\displaystyle =\frac12\rho_1
\int\frac{\partial}{\partial\beta_1}\left[\beta_1d^3\phi(r_1d)\right]
n_1(r_1){{\mathrm d}}\vec r_1\\
\displaystyle &\displaystyle =\frac12\rho
\int\left[\phi_1(r_1)
+ A_1 (\phi_1(r_1)
+\frac13d^3
\frac{\partial\phi(r_1d)}{\partial d})
\right]
n_1(r_1){{\mathrm d}}\vec r_1\\
\displaystyle &\displaystyle =\frac12\rho
\int\left[\phi_1(r_1)
+ \frac13 A_1 \nabla(\vec r_1\phi_1(r_1))\right]
n_1(r_1){{\mathrm d}}\vec r_1\\
\displaystyle &\displaystyle =u_1 + A_1\,v_1,
\end{array}
\end{displaymath}
in accordance with the right hand side of Eq~(\ref{eq:scoza:transformed}).

%77777777777777777777777777777777777777777777777777777777777777777777777777
\section{Conclusion}
 
We have investigated a method to perform SCOZA evaluations for
realistic molecules with soft cores. A temperature dependent effective
hard core diameter is then introduced. For thermodynamic
self-consistency it turns out that the excess internal energy of the
reference system is important and should be treated accurately. For
this purpose the contact value of the pair correlation function at the
hard sphere surface is needed. Since SCOZA does not
give such a reliable contact value, it is instead obtained by use of
the virial theorem. In section~\ref{sec:unit} the SCOZA problem with
varying $d$ is transformed to a simpler one with fixed $d=1$. 
This transformation not only eliminates some of the difficulties
associated with temperature-dependent diameter but also provides an
independent method of evaluation that can be useful as a test of the
internal consistency of results. 

The results of this work have been obtained along with numerical work
to implement and solve new problems when soft cores are considered. So
far we have only considered simple test functions $d(\beta)$ to obtain
a program that can handle varying $d$ when using analytic expressions.
In this respect the transformation to fixed $d=1$ in
section~\ref{sec:unit} has been verified numerically by solving SCOZA
with both the original and transformed equations by putting for
simplicity the virial type integral for $v$ and thus the one for $v_1$
equal to zero.  For explicit evaluations with the LJ interaction it is
necessary to evaluate integral~(\ref{eq:d:def}) for $d(\beta)$ and the
less trivial one for $v$; preliminary computations already including
$v$ are encouraging.  Furthermore, it will then also be desirable to
study the influence of different prescriptions for $d(\beta)$ and for
the splitting of the potential into a reference part and a perturbing
attractive part.  We intend to extend our computations in this way.

\section*{Acknowledgments}

AR\ gratefully acknowledges financial support from \textit{Fonds zur
F\"orderung der wissenschaftlichen Forschung (FWF)} under
project~J2380-N08.

\end{document}